\documentclass[onecolumn,showpacs,preprintnumbers,amsmath,amssymb,superscriptaddress]{revtex4}

\usepackage{graphicx}
\usepackage{dcolumn}
\usepackage{bm}
\usepackage{longtable}

\begin{document}
\title{Random walks interacting with evolving energy landscapes}

\author{E. Agliari}
\affiliation{Dipartimento di Fisica, Universit\`a degli Studi di Parma, Viale Usberti 7/a, 43100 Parma, Italy} 
\author{R. Burioni}
\affiliation{Dipartimento di Fisica, Universit\`a degli Studi di Parma, Viale Usberti 7/a, 43100 Parma, Italy}
\author{D. Cassi}
\affiliation{Dipartimento di Fisica, Universit\`a degli Studi di Parma, Viale Usberti 7/a, 43100 Parma, Italy}
\author{A. Vezzani}
\affiliation{CNR- INFM Gruppo Collegato di Parma, Viale Usberti 7/a,
43100 Parma, Italy}
%

%
\date{\today}

\begin{abstract}We introduce a diffusion model for energetically
inhomogeneous systems. A random walker moves on a spin-S Ising
configuration, which generates the energy landscape on the lattice
through the nearest-neighbors interaction. The underlying
energetic environment is also made dynamic by properly coupling
the walker with the spin lattice. In fact, while the walker hops
across nearest-neighbor sites, it can flip the pertaining spins,
realizing a diffusive dynamics for the Ising system. As a result,
the walk is biased towards high energy regions, namely the
boundaries between clusters. Besides, the coupling introduced
involves, with respect the ordinary diffusion laws, interesting
corrections depending on either the temperature and the spin
magnitude. In particular, they provide a further signature of the
phase-transition occurring on the magnetic lattice.
\end{abstract} 
\maketitle
\section{Introduction}
\label{intro} The problem of diffusion on inhomogeneous media is
attracting much attention, due to its fundamental importance in
nearly every field of science and engineering
\cite{avramov,bricmont,nitzan,bustingorry,harrison,druger,masuda}.
In fact, for a proper description of many diffusive systems (for
example proteins ``sliding'' on DNA, diffusion of charge carriers
in solids, flow through porous media, etc), apart from the motion
of the particles, also the underlying environment must be included
\cite{slutsky,bookout}.

There are several ways to introduce disorder or, more generally
inhomogeneity. For example, it can be geometric, due to an
irregular lattice structure, or energetic. In the latter case, the
lattice sites (or bonds) are assigned different energy states and,
consequently, the walker is biased towards sites corresponding to
potential wells (or small energy barriers). Moreover, disorder can
be deterministic or random and it can be dynamic (the environment
is renewed at each jump of the walker) \cite{avramov,levitsky}, or
static (the environment is frozen in a particular configuration)
\cite{arapaki,haus}. Particles diffusing on such structures can
also be endowed with memory effects \cite{schulz1,schulz2}, or be
influenced by the distribution of other diffusing particles on the
same structure \cite{schulz3}.

In our model, an evolving, inhomogeneous energy landscape is
introduced, by coupling the random walk with a spin-S Ising
system. More precisely, we assume a spin-S arbitrary lattice and
we let a random walker moving on it. The relevant energy landscape
is then obtained by relating each lattice site with the pertaining
nearest-neighbor interaction according to the Ising Hamiltonian.
In other words, the walker moves on a lattice where each site is
occupied by a spin $\sigma_i \in [-S, S]$ and which generates the
energy environment through the Ising interaction.
Now, if we make the Ising ferromagnet be in contact with a
heat-bath, by varying the temperature parameter the spin
configuration evolves and then, also the energy landscape is
modified. In particular, the temperature acts as a dispersion
parameter \cite{sapag}, being able to control the roughness of the
energetic environment. In fact, when the temperature is
sufficiently low, the lattice is ferromagnetic and the energy
landscape is flat, viceversa when $T \rightarrow \infty$ the
energy landscape is rugged.

However, differently from the dynamic, inhomogeneous systems
introduced in previous works \cite{avramov,levitsky}, where the
energy landscape was updated from external forces, here we assume
that the random walker, while hopping across the sites of the
underlying lattice, flips the relevant spins. Hence, during the
diffusion of the walker on the lattice the magnetization and the
energy properly vary. In fact, as we will see later, by defining a
suitable spin-flip probability, the random walker is able to
provide a diffusive thermal dynamics \cite{earlier,buonsante}. In
particular, as a result of our assumptions, the walker is now
biased towards such sites that, by flipping the relevant spin, an
energy gain can be achieved.

Therefore, the problem of the RW on an inhomogeneous energy
landscape is non trivially extended to the problem of their
interaction: the RW affects, and is biased by, the energy
landscape. In other words, there are two interplaying stochastic
processes: the motion of the walker and the evolution of the spin
configuration.

Our work will be mainly numerical and the algorithm implemented is
very general, being easily appliable to arbitrary lattices, made
up of spins which can assume an arbitrary, finite number of
states.

The aim of this work is then to characterize the random walk
introduced, especially highlighting how its interaction with the
magnetic lattice affects its diffusion. In particular, it would
turn out to be interesting to relate the behavior of the walker
with the evolution of the energy landscape, namely with the
evolution of the magnetic lattice. Hence, we analyze our biased
random walker (BRW) at different temperatures and then we compare
results with those, already known, relevant to the ordinary,
unbiased random walker (URW). Interestingly, as we will show,
though their asymptotic behaviors agree
\cite{montroll,weiss2,weiss}, temperature dependent corrections
have to be introduced. In particular, the functional laws
describing the behavior of our BRW are URW-consistent, while the
pertinent multiplicative factors peak at $T_c$. Therefore, effects
due to the coupling between the walker and the magnetic system are
strongest as the latter undergoes its phase transition. In other
words, the diffusion of the BRW provides signatures of the phase
transition occurring on the magnetic lattice. Besides, in order to
understand to what extent the spin magnitude influences the walker
diffusion, we take into account both spin-1/2 and spin-1 Ising
systems.

Finally, notice that all the measures that are being explained are
performed after the magnetic system has reached a steady state.

The layout of the paper is as follows. In Sec.~\ref{dyn} we
explain how the energy landscape is generated and how our BRW can
update it; we also underlines the differences with respect to the
URW. In Sec.~\ref{B_U} we show how, under some conditions, such
differences can vanish and then the URW is recovered. In Sections
\ref{Lattice}-\ref{Numbers} we describe the numerical simulations
performed, useful to characterize the walker behavior. We
especially analyze in details the covering time, the number of
returns to the origin and of distinct sites visited since they
better emphasize the relationship between the walker and the
magnetic lattice. Finally, Sec.~\ref{concl} contains a summary and
a discussion of results.

\section{Diffusive Dynamics}
\label{dyn} In this work we deal with a RW  moving on, and
interacting with the energy landscape generated by the following
Hamiltonian applied to the magnetic configuration of a spin-S
Ising system:
\begin{equation} \label{eq:hamiltonian}
{\cal H} = - \frac{J}{S^2} \sum_{i,j}^{N} A_{ij} \, \sigma_i
\sigma_j + \frac{h}{S}\sum_{i}^{N} \sigma_i.
\end{equation}
The spin variable $\sigma$ may take the (2S+1) values $-S, -S+1,
{\ldots}, S-1, S$ and $A{ij}$ is the adjacency matrix associated
to the arbitrary network where spins are placed on. Hence, the
first sum only involves nearest neighbor pairs, according to the
chemical distance.

Though our analysis has been performed on a toroidal squared
lattice with $J = 1$, $h = 0$ in order to focus the attention on
the very dynamical effects, in the remaining of this section we
make assumptions on neither the structure of the lattice nor on
the spin magnitude (though finite).

Hitherto we have just explained how the energy landscape is
generated starting from a discrete spin configuration, while now
we will describe how the coupling between the magnetic lattice and
the walker works.

The random walker is assumed to be able to move on
nearest-neighbor sites or stop, and it can also flip the spin
pertaining to the reached site (notice that, when the spin
magnitude is very large, the latter procedure can be quite complex
due to a $(2S+1)$-manifold choice). Therefore, our model displays
two interplaying stochastic processes: the diffusion of the walker
on the lattice and the evolution of the spin configuration. Such
processes can be considered consequentially (firstly decide the
site to move towards and then select the relevant spin state or
viceversa) or contemporary (consider all possible combinations
spin+site and choose one of them). Indeed, in any case, there
exist many different ways to rule this system, ranging from
completely random to completely deterministic.

The assumptions for our model have been taken in order to realize
a proper diffusive dynamics for the Ising model. Such a dynamics
was introduced in a previous paper \cite{earlier} where the
thermodynamics aspects were investigated. In particular, it was
found that our diffusive thermal dynamics is actually able to
drive the system towards a non canonical equilibrium state, which
depends on the temperature but not on the particular initial spin
configuration. As far the critical behavior, it preserves the
universality class, though the critical temperature is increased:
\begin{eqnarray}
T_{c}^{S=1/2} = 2.602(1)\label{eq:Tc_Mezzo}
\\
T_{c}^{S=1}=1.955(2)\label{eq:Tc} .
\end{eqnarray}
Because of this sort of right-shift with respect the canonical
dynamics, it is worth underlining that, in the remaining of the
paper, when we refer to the critical range or temperature, it is
always meant according to the diffusive dynamics.

Now, let us see in detail how the probability running our RW is
defined. First of all, it contemporary takes into account the
motion of the walker and the spin-flip procedure, besides, it is
local since it only depends on the magnetic configuration of RW's
nearest-neighbor sites. More precisely:
\begin{equation} \label{eq:probability}
{\cal P}_T(\vec{s},j | \vec{s_0},i) = \frac{p_T(\vec{s},j)
(A_{ij}+ \delta_{ij})}{\sum_{\{\vec{s'}\}} \sum_{j=0}^{z_i}
p_T(\vec{s'},j)}.
\end{equation}
represents the probability that the walker, being on site $i$ with
coordination number $z_i$, jumps on a n.n. site $j$ and realizes
the magnetic configuration $\vec{s}$. The spin configuration
before the jump is denoted as $\vec{s_0}$, while $\{\vec{s'}\}$ is
the set of the new possible configurations. Furthermore,
\begin{equation} \label{eqn:glauber}
p_T(j,\vec{s})= \frac{1}{1+e^{[\beta \Delta E_j(\textbf{s})]}}
\end{equation}
is derived from the usual Glauber probability (see \cite{earlier}
for more details). Also,
\begin{equation} \label{eq:delta_e}
\Delta E_j(\vec{s}) = (\sigma_j^i-\sigma_j^f)\sum_{j\sim
k}\sigma_k,
\end{equation}
is the energy variation consequent to the process, where
$\sigma_j^i$ and $\sigma_j^f$ represent the spin-state on site $j$
before and after the flip procedure, respectively.

You can notice that, at each step, the walker can choose among
$z_i + 1$ sites to move towards (or stay on) and, contemporary, it
can also choose if flip the relevant spin, being biased in order
to achieve an energy gain.

Hence, all in all, there are $(z_i+1)\times(2S+1)$ options
including that the magnetic configuration of the system, as well
as the position of the walker, will possibly remain unchanged.

Of course, in a $d$-dimensional hypercubic lattice, the number of
nearest-neighbors does not depend on the particular site and
$z_i=2d, \forall i$.

Notice that, the hopping rate between two sites is, in general,
different going forward and backwards, so that the random walk is
asymmetric. Consequently, as stressed in \cite{earlier}, this kind
of dynamics does violate the detailed balance condition and the
equilibrium states achieved are non-canonical.

In traditional models of diffusion on energetic landscapes, the
jump rate is typically controlled by the local energy at the start
point or by the energy-barrier height between start and end points
\cite{haus}. Though previous equations
imply that
spin-flips occur on the site where the walker is moving towards,
we can as well think our model in terms of energetic barriers. In
fact, energy-barriers are lower for nearest-neighbor sites which
let, by means of spin-flip, a higher energy gain. Furthermore,
such behavior of the walker is consistent with the physical
systems which have inspired the model \cite{earlier,buonsante}.

It is now worth comparing our RW with the traditional unbiased
random walker, usually defined according to the probability:
${\cal P}(i,j)=\frac{A_{ij}}{z_i}$, since, at every step the
walker must move and the hopping probabilities are isotropic and
do not depend on time. In this work, in order to establish a
stronger analogy with our BRW, we will endow the unbiased random
walker with a waiting probability so that, from site $i$, the
possible, equivalent, choices are $z_i+1$. In other words, we
allow repetitions within the succession defining the trajectory of
the walker and
\begin{equation} \label{eq:probability_URW}
{\cal P}(i,j)=\frac{A_{i,j}+ \delta_{i,j}}{z_i+1},
\end{equation}
where $j$ is a nearest neighbor of $i$'s or, possibly, the site
$i$ itself. In the following, we will refer to this able-to-stop
unbiased random walker as SURW.

It is known \cite{burioni} that, for the URW, the possibility of
staying on the same site is crucial in the short time regime,
while in the long time behavior it has no important consequences.
As we will see later, an analogous long-time effect is also
experienced by our BRW.

\section{BRW recovers URW}
\label{B_U} As mentioned in Section \ref{intro}, the temperature
parameter can tune the roughness of the energy landscape. In
particular, when $T$ is sufficiently low, the magnetic lattice is
homogeneous, the energy landscape is flat and we expect to recover
the URW case. On the other hand, when $T \rightarrow \infty$, a
completely disordered lattice and, consequently, a rugged energy
environment, is achieved. Nevertheless, since the energy
variations consequent to whatever possible spin-flip would be very
small compared with $\beta$, we again expect to recover the URW
case. Then, in this section, we want to prove that
Eq.~(\ref{eq:probability}) recovers
Eq.~(\ref{eq:probability_URW}), under the conditions $\beta
\rightarrow \infty$ and $| \displaystyle \sum_{i=0}^{N} \sigma_i|
= NS$, or $\beta \rightarrow 0$. Firstly, let us consider the
former case with $\sigma_i=S, \forall i$. Suppose that the walker
jumps from site $i$ to $j$, with coordination numbers $z_i$ and
$z_j$ respectively and that, consequently, $\sigma_j$ is flipped
in $\sigma ^\prime$. Then, Eq.~(\ref{eq:probability}) can be
rewritten as
\begin{equation} \label{eq:probability_rewritten}
{\cal P}(i,j,\sigma_j= \sigma ^{\prime})=\frac{[1+E^{z_j(S -
\sigma ^\prime)}]^{-1}}{\displaystyle \sum_{k=0}^{z_i}
\displaystyle \sum_{l=0}^{2S} \frac{1}{1+E^{z_k l}}},
\end{equation}
where $E=e^{J\beta/S}$ and $i$'s nearest-neighbors have been
numbered from $0$ (the walker remains on $i$) to $z_i$. We also
dropped the factor $(A_{ij}+ \delta_{ij})$, because we assume $j$
to be linked to $i$, or, possibly, $i=j$. Now, since $\beta
\rightarrow \infty$, then  $E \rightarrow \infty$ and we can
write:
\begin{equation} \label{eq:probability_recovered}
{\cal P}(i,j,\sigma_j = \sigma^{\prime}) = \left\{
\begin{array}{rl}
\frac{1}{z_i+1}+ O(E^{-\zeta}) & \mbox{if } \sigma^\prime = S \\
O(E^{-z_j(S- \sigma ^\prime)}) & \mbox{if } \sigma^\prime \neq S,
\end{array}
\right.
\end{equation}
where $$\zeta = \min_{k=0,...,z_i}(z_k)$$

Conversely, when $\beta \rightarrow 0$, then $E \rightarrow 1$,
and, recalling that $S$ is finite, you can easily find that:
\begin{equation} \label{eq:probability_recovered_2}
{\cal P}(i,j,\sigma_j = \sigma^{\prime})= \frac{1}{(z_i+1)(2S+1)}
+ O(\beta J \xi),
\end{equation}
where $$\xi = \max_{k=0,...,z_i}(z_k).$$ The previous equation
depends neither on $\sigma^{\prime}$, nor on the magnetic
configuration of the lattice and hence, all in all, the
probability of jumping from a site to another recovers
Eq.~(\ref{eq:probability_URW}).

In the following sections we analyze the behavior of the walker
introduced, focusing the attention on those aspects which are
mostly affected by its interaction with the magnetic lattice.
Results will be further stressed by comparison with their SURW
counterparts.

\section{Visit Lattice}
\label{Lattice} 
\begin{figure}[tb]
\resizebox{0.45\columnwidth}{!}{\includegraphics{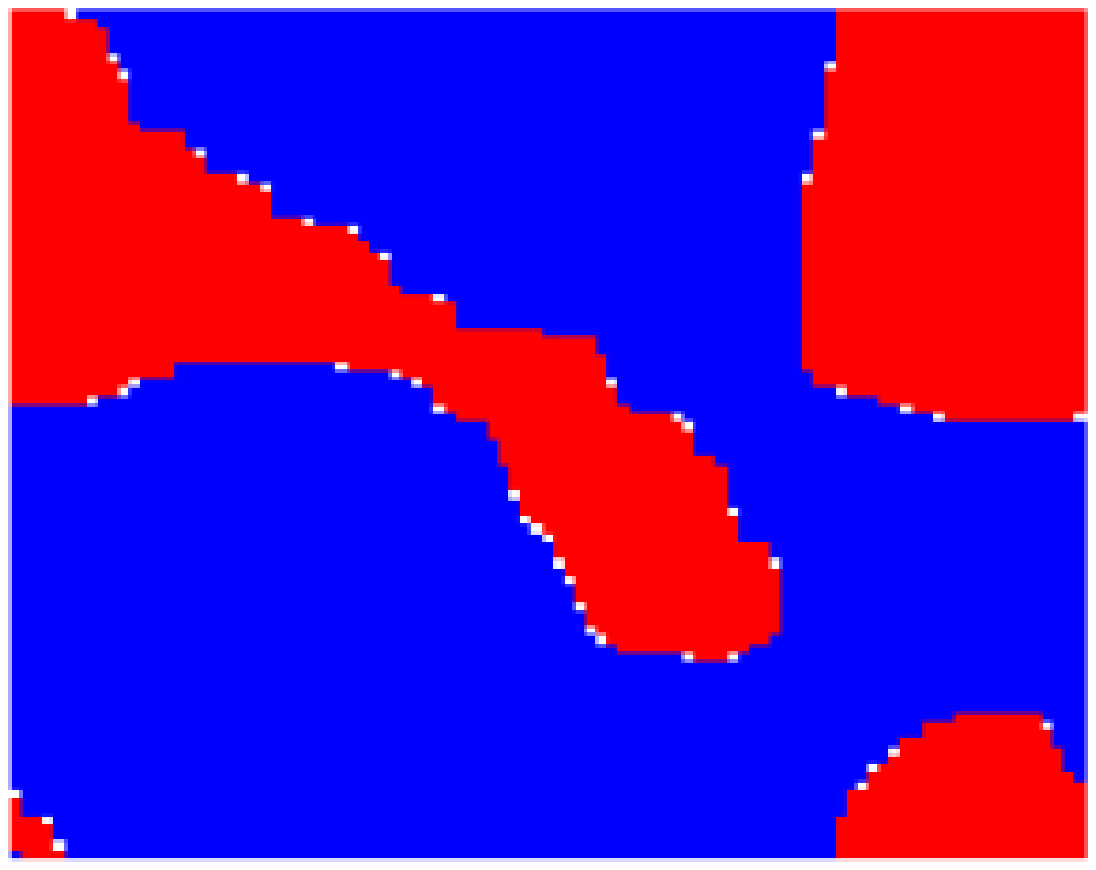}}
\resizebox{0.45\columnwidth}{!}{\includegraphics{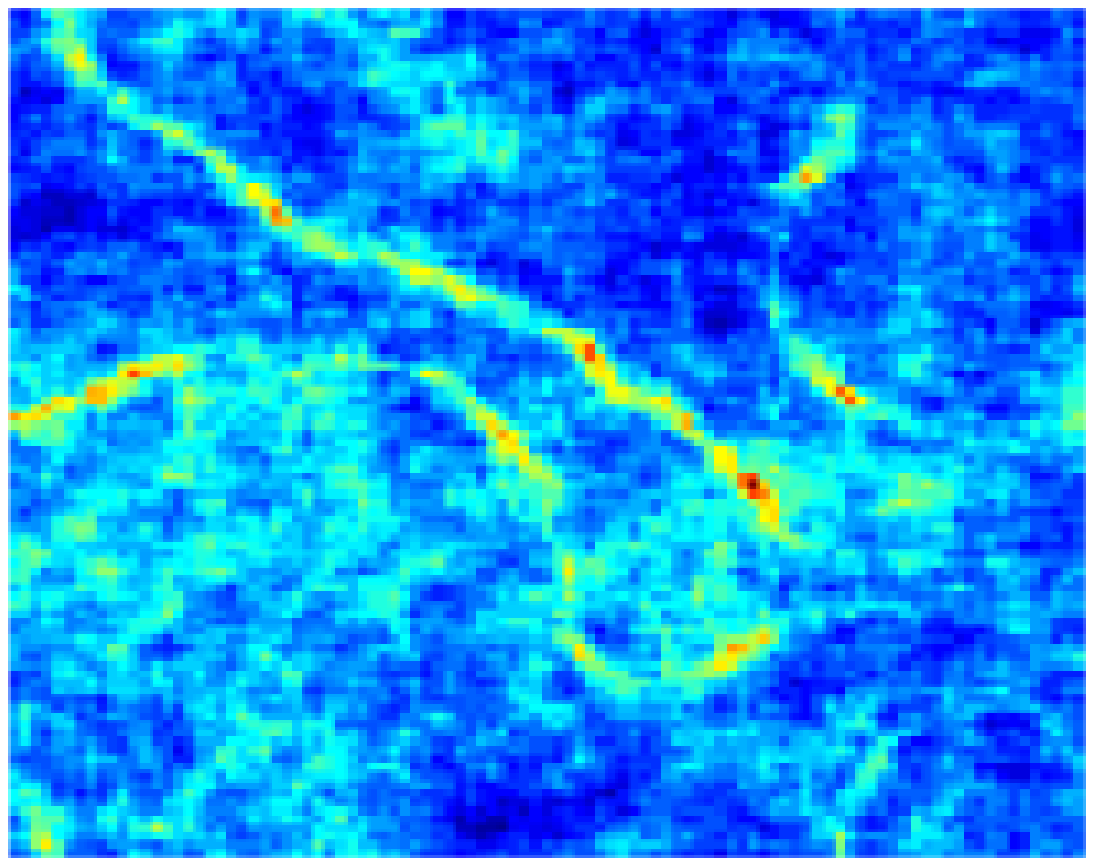}}
\caption{(Color online) Magnetic (left panel) and Visit (right
panel) lattice for a $100 \times 100$ spin-1 Ising system subject
to the diffusive dynamics described in Sec.~\ref{dyn}. The sample,
initially paramagnetic, was suddenly cooled at low temperatures
($T < T_c$). Note that null spins (white) arrange themselves just
on those sites such that their n.n. provide, all in all, a neutral
magnetization; these sites usually belong to cluster boundaries.
In the left panel the warmest colors are reserved to the most
visited sites which are the ones corresponding to boundaries
between clusters; analogous results hold for
spin-1/2.}\label{fig:boundary_magrw}
\end{figure}
In this work a task of ours is to relate the
motion of the walker with the magnetic configuration of the Ising
lattice representing the energy landscape. To this aim, we
introduce the visit lattice, meant as the $L \times L$ array whose
elements are incremented by a unit each time the walker passes
through the pertaining site. In Fig.~\ref{fig:boundary_magrw} such
a lattice is compared with the magnetic one. From
Eq.~(\ref{eq:probability}) we expect the walker to be attracted
towards high energy regions which, in our model, corresponds to
borders between clusters. Of course, this attraction would affect
the distribution of visit numbers on the lattice provided that the
parameter $\beta$ is not so small to make any spin-flip equally
probable (see Eq.~(\ref{eq:probability_recovered_2})). Actually,
in Fig.~\ref{fig:boundary_magrw}, the attraction felt by the
walker is strong enough to generate detectable effects on the
visit lattice: as expected, the most ``popular sites'' are just
those belonging to the perimeter of cluster. As a consequence, the
visit lattice mirrors the magnetic lattice: looking at the former
one can derive the spin configuration and vice versa.

\section{Local Energy}
\label{Local}

\begin{figure}[tb]
\resizebox{0.6\columnwidth}{!}{\includegraphics{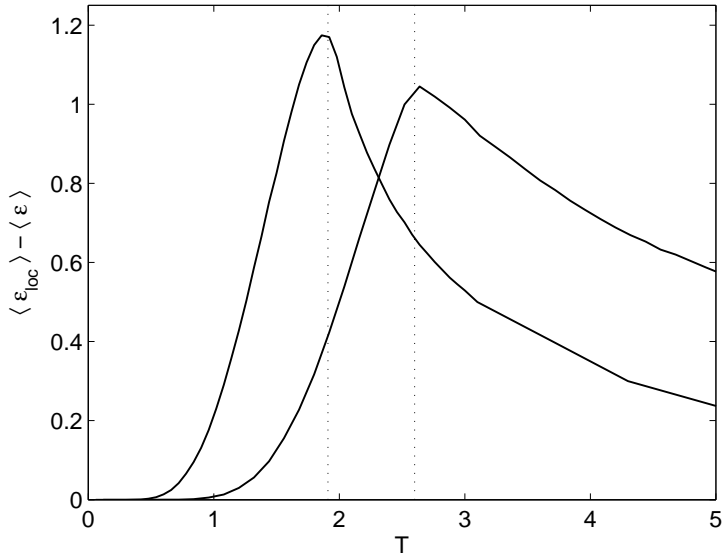}}
\caption{$\langle \epsilon_{loc} \rangle - \langle \epsilon
\rangle$ for a $240 \times 240$ spin-1/2 and spin-1 Ising systems
initialized ferromagnetic and then heated. The two plots can be
easily distinguished, as they peak at the relevant critical
temperature (dashed line) and the former case shows a higher value
of $T_c$.}\label{fig:corr}
\end{figure}

In the previous section we showed that, according to
Eq.~\ref{eq:probability}, BRW does not move freely, but it can be
forced to stay nearby high energy regions. Hence, we expect the
local energy $\epsilon_{loc}$ to be larger than the energy of the
whole system $\epsilon$. Now, we wonder if the difference between
such quantities is somehow temperature dependent. Therefore, we
consider the quantity $\tilde{\epsilon} = \langle \epsilon_{loc}
\rangle - \langle \epsilon \rangle$, where, we recall,
$\epsilon_{loc}$ represents the energy relevant to the site $i$
occupied by the walker, namely $\epsilon_{loc} = \sigma_i
\displaystyle \sum_{i \sim j} \sigma_j$. As shown in
Fig.~\ref{fig:corr}, as long as the temperature is sufficiently
low, the lattice appears homogeneous and $\tilde{\epsilon}$ is
null. However, heating the sample, some domains develop and, since
the walker verges on their borders, $\langle \epsilon_{loc}
\rangle$ can increase more than $\langle \epsilon \rangle$ so that
$\tilde{\epsilon}$ rises. While approaching the critical
temperature, more and more clusters arise and the walker is more
and more likely to be found on their boundaries, which explains
the maximum in $T_c$. Conversely, at high temperature, when the
paramagnetic phase has been reached, $\langle \epsilon \rangle$
gets to $\langle \epsilon_{loc} \rangle$.

Note also that, in Fig.~\ref{fig:corr}, the peak relevant to the
spin-1 case is sharper, which means a stronger interaction between
the walker and the magnetic lattice.

We also measure $\tilde{\epsilon}$ for an unbiased random walker
allowed to rest and moving on an Ising lattice subject to a
non-diffusive dynamics. Of course, in this situation, the walker
is completely useless for the evolution of the system,
nevertheless its behavior underlines that results obtained for the
BRW are really due to its interaction with the magnetic system. In
fact, we find that, for the SURW, $\tilde{\epsilon}$ remains close
to zero without displaying any significant dependence on the
temperature.

Therefore $\tilde{\epsilon}$ provides a signature that, as far our
BRW, at the critical temperature interesting phenomena occur, not
only in thermodynamics terms.

\section{Covering Time}
\label{Covering}
\begin{figure}
\resizebox{0.45\columnwidth}{!}{\includegraphics{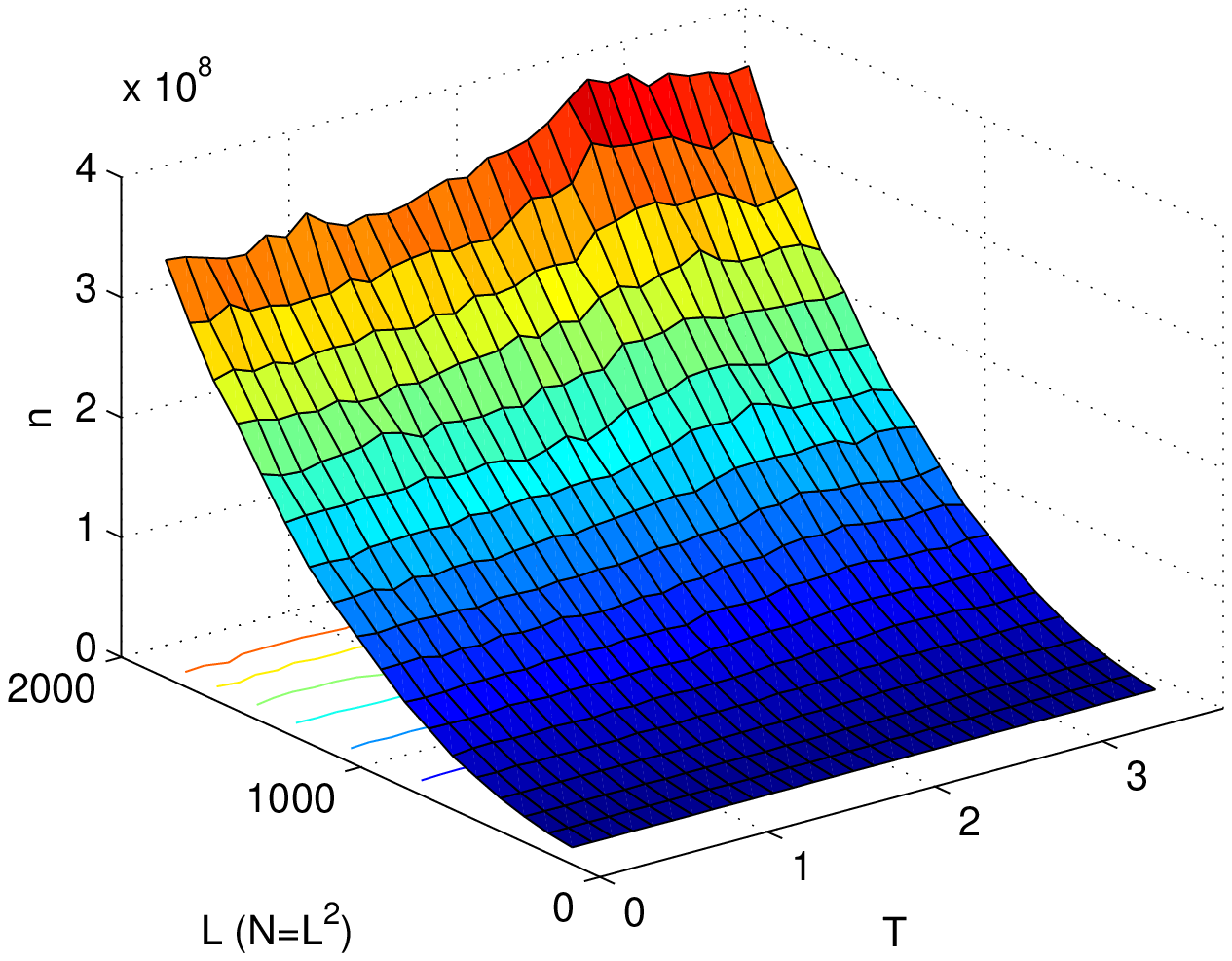}}
\resizebox{0.45\columnwidth}{!}{\includegraphics{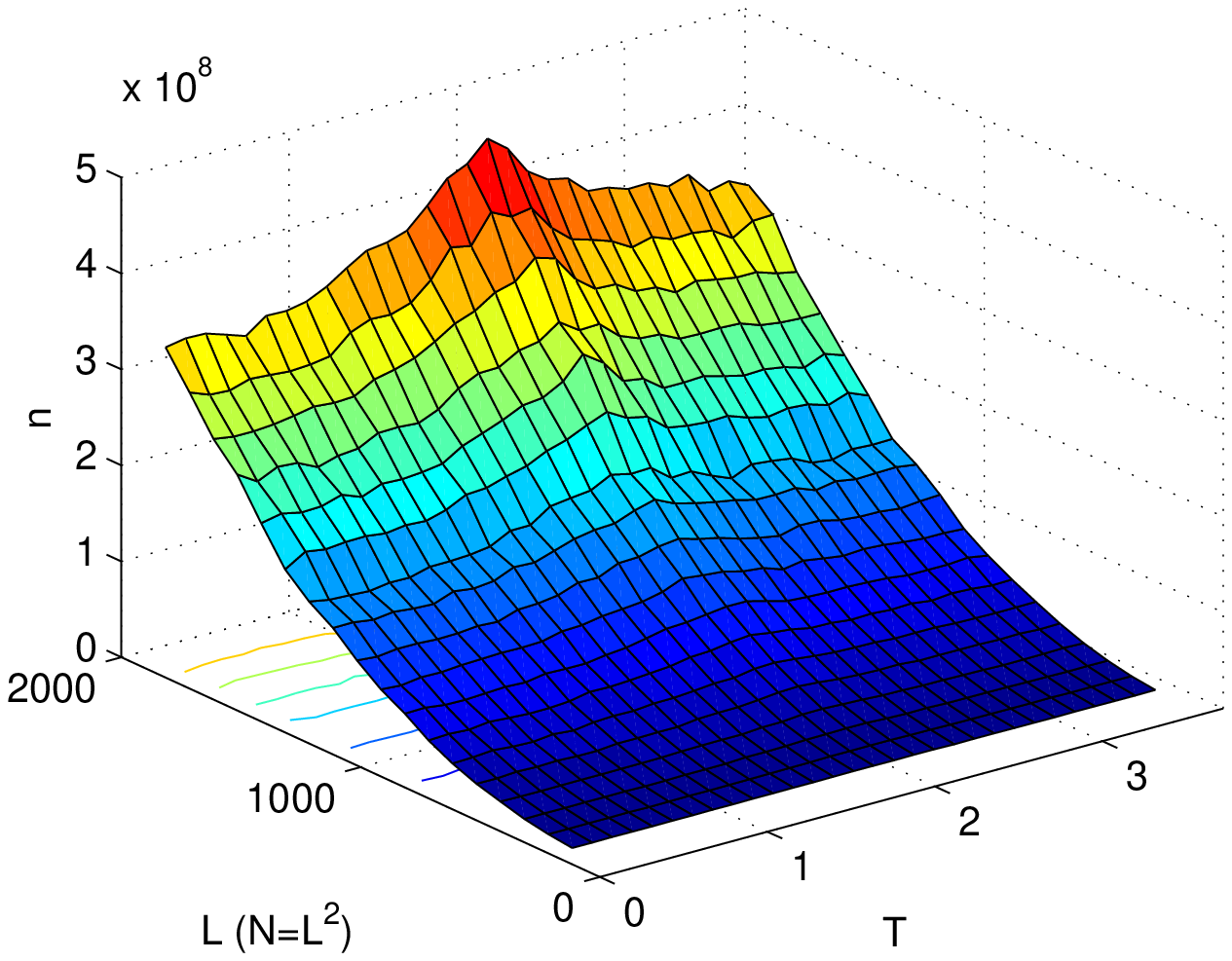}}
\caption{(Color online) \label{fig:CovTime3} Covering time versus
temperature and size of the lattice for the biased random walker
moving on a spin-1/2 (left panel) and spin-1 (right panel) Ising
system. In both cases the outline depends strongly on the size and
the temperature affects the covering time just during the critical
regime.}
\end{figure}
\begin{figure}
\resizebox{0.6\columnwidth}{!}{\includegraphics{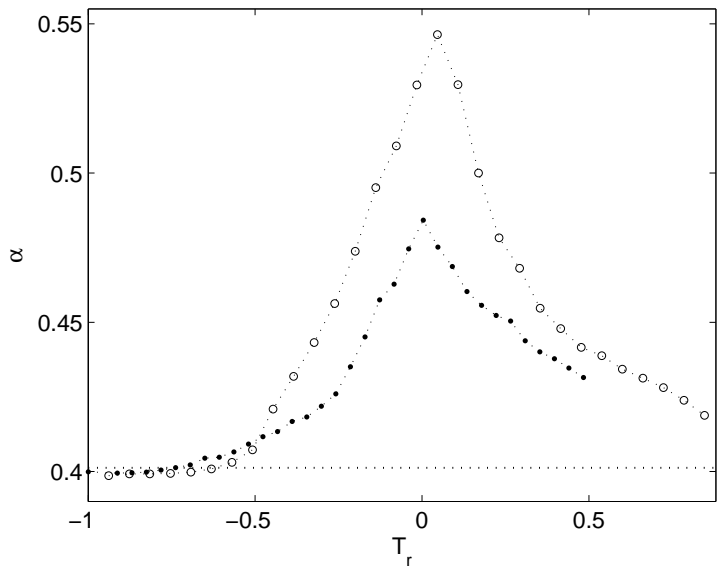}}
\caption{\label{fig:Alpha_Compare_I} Multiplicative constant
$\alpha$ versus reduced temperature for the BRW applied to the
spin-1/2 ($\bullet$) and spin-1 ($\circ$) Ising systems. Both
functions peaks at zero, while $\alpha_{URW}$ (dashed line) is
temperature independent. Note that (within the error)
$\alpha_{SURW} \leq \alpha_{BRW}^{S=1/2} \leq \alpha_{BRW}^{S=1}
$. }
\end{figure}
The previous two sections pointed out that the hopping-flipping
probability, defined in Section \ref{intro}, actually biases the
walker towards high energy regions and that, in the critical
range, the coupling between the walker and the magnetic lattice is
even more important. Hence, the phase transition also emerges from
the behavior of the walker; this interesting feature will be
especially taken into account in the following analysis. In
particular, we now consider the covering time, namely the time (in
unit step) taken by the walker to visit all $N$ sites making up
the lattice. We recall that the lattice is squared and endowed with periodic
boundary conditions, so that the walker can actually cover an
infinite distance on it.

As depicted in Fig.~\ref{fig:CovTime3}, for both spin-1/2 and
spin-1 systems, the covering time ${\cal T}_N$ measured for the
BRW increases with the size of the lattice and a temperature
dependence is also noticeable. In particular, there is an increase
in the covering time at about the critical temperature, which has
been previously measured \cite{earlier} revealing to be fairly
larger than the canonical one (Eqs.~\ref{eq:Tc_Mezzo} and
\ref{eq:Tc})

As far the dependence on the total number of sites $N$, it is
consistent with the logarithmic law:
\begin{equation} \label{eq:covering}
{\cal T}_N = \alpha N (log N)^2
\end{equation}
found by analytical \cite{jonasson,dembo} as well as numerical
\cite{nemirovsky} methods applied to an unbiased random walk.
Then, the very effects due to the bias have to be tracked down in
the multiplicative factor $\alpha$. In fact, by fitting, according
to Eq.~(\ref{eq:covering}), the data relevant to both spin
systems, we evidenced that $\alpha_{BRW}^{S=1/2}$ and
$\alpha_{BRW}^{S=1}$ depend on $T$ and are larger than
$\alpha_{URW}$ found in \cite{jonasson,nemirovsky}. Notice that
either the possibility of maintaining the same position and the
interaction with the magnetic lattice concur in lengthening the
covering time, but the role played by the latter is non trivial.
In particular, it makes $\alpha_{BRW}^{S=1/2}$ and
$\alpha_{BRW}^{S=1}$ exhibit a maximum at about $T_c$, namely, in
the critical region, it takes more time the BRW to cover the
lattice. On the other hand, the SURW displays a covering time
still consistent with Eq.~(\ref{eq:covering}) but, of course,
independent on $T$. Note that, as shown in
Fig.~\ref{fig:Alpha_Compare_I}, within the error ($<2 \%$),
\begin{equation}
\label{eq:alphas} \alpha_{SURW} \leq \alpha_{BRW}^{S=1/2} \leq
\alpha_{BRW}^{S=1}.
\end{equation}
As expected, the BRW, with respect the SURW, is slowed down since
it may be  ``trapped'' nearby high energy regions constituted by
sites where several, energetically favorable, sin-flip are
possible. On the other hand, the quantities in
Eq.~(\ref{eq:alphas}) are all comparable at low temperatures. In
fact, during the ferromagnetic phase, when $T \ll T_c$, the
lattice appears homogeneous and the bias has no effect; an
analogous phenomenon is expected at very high temperatures (in
section \ref{B_U} we proved that, indeed, in these cases the BRW
recovers the SURW). Besides, since in the spin-1 case the walker
has to manage a greater number of possibilities, the slowing down
effect is even higher and the relevant peak in
Fig.~\ref{fig:Alpha_Compare_I} is more marked.

Note that, with a non-diffusive dynamics, the time required to
scan each lattice site at least once can be much smaller. For
example, by adopting the type-writer sequence, ${\cal T}_N$ is
reduced by a factor $4 \alpha_{BRW} \, (log L)^2$. For this
reason, our diffusive dynamics may be though as ``slow''.

\section{Returns to the origin and distinct sites visited}
\label{Numbers}
In this Section, we want to deal with other two characteristic
quantities concerning the BRW: the number of returns to the origin
$R_n$ and of distinct sites visited $D_n$, after an $n$-step walk.
Actually, we ought to distinguish between two regimes:
\begin{enumerate}
\item if $n$ is a step number satisfying $n \ll N$, since the
walker will not have sampled a substantial number of sites of the
lattice, the lattice will appear to be infinite; \item in the long
time ($N \ll n$) the walker will appreciate the toroidal effect
due to the periodic boundary conditions.
\end{enumerate}
Therefore, according to the walk-length, our results will be
compared with those analytically known and relevant to the URW on
an infinite or periodic 2-dimensional lattice, respectively. As we
will see, in both ranges, the exponents found for $R_n$ and $D_n$
agree with known results, while, as far the multiplicative
factors, one has to introduce a dependence on $T$.

\begin{figure}
\resizebox{0.45\columnwidth}{!}{\includegraphics{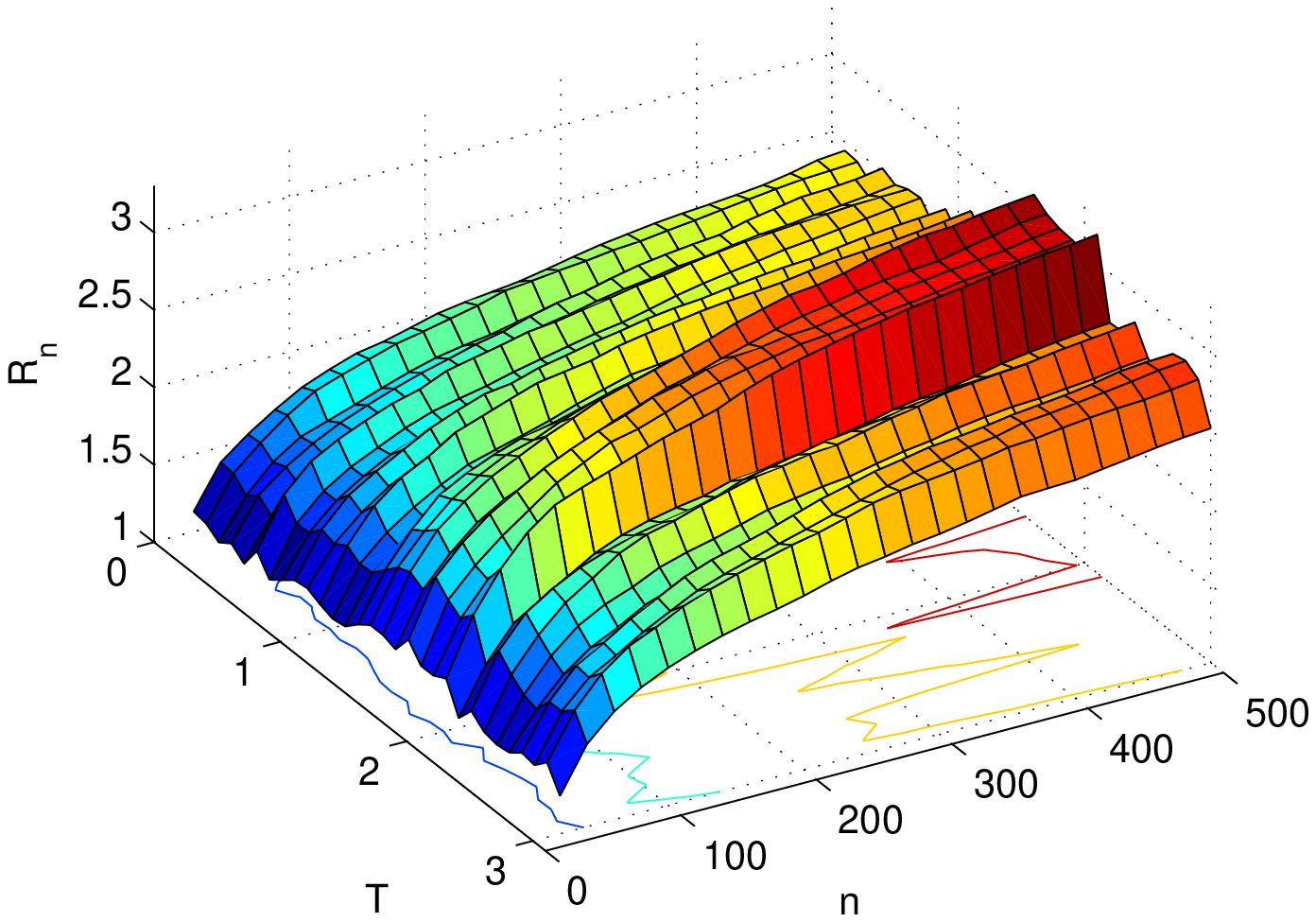}}
\resizebox{0.45\columnwidth}{!}{\includegraphics{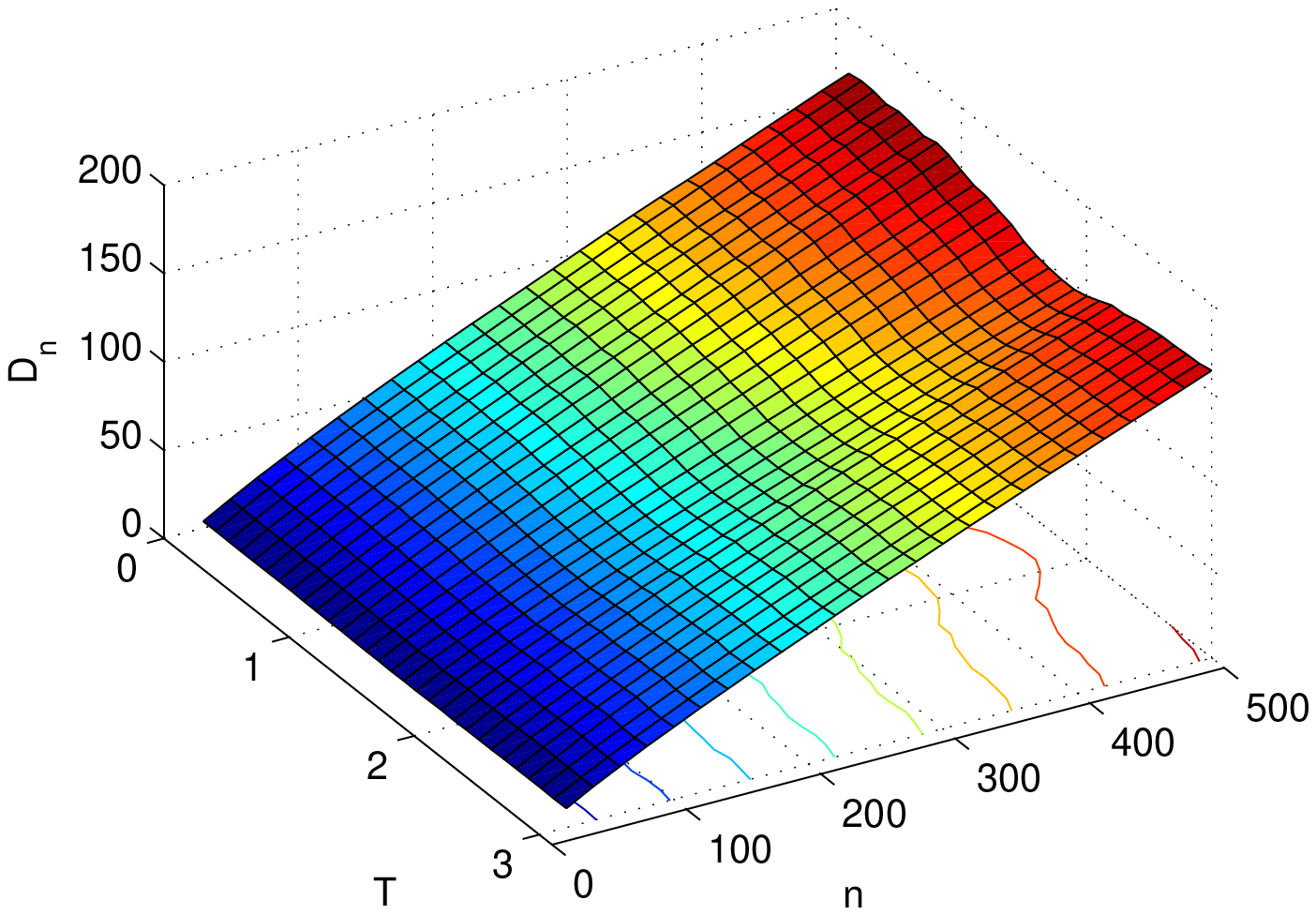}}
\caption{(Color online) \label{fig:Origin_Distinct_Short} Average
number of returns to the origin (left panel) and of distinct sites
visited (right panel) versus temperature and number of steps made
by the BRW on a $240 \times 240$ spin-1 Ising lattice. Analogous
results were found for the spin-1/2 system, though the critical
phenomena are less emphasized. Note that, as $n \ll N$, the walker
can not realize the finiteness of the lattice.}
\end{figure}
\begin{figure}
\resizebox{0.6\columnwidth}{!}{\includegraphics{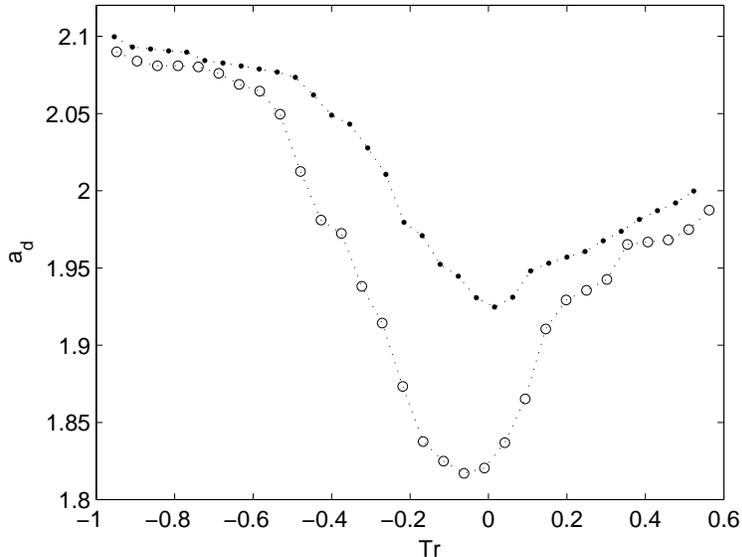}}
\caption{\label{fig:Fit_Distinct_Short} Fit parameter $a_D$ versus
reduced temperature relevant to the BRW on a $240 \times 240$
spin-1/2 and spin-1 Ising lattice. Both functions exhibit a
minimum at about the relevant critical temperature ($T_c \approx
2.6$ and $T_c \approx 1.96$ respectively) which is deeper in the
latter case. Conversely, at low temperatures, plots overlap.}
\end{figure}

Let us firstly consider the short time case. Known results
\cite{montroll,weiss2} about the average number of returns to the
origin $R_n$ and the number of distinct sites visited $D_n$, after
an $n$-step walk for an URW, state that:
\begin{equation} \label{eq:origin}
R_n \sim a_R \: log n,
\end{equation}
\begin{equation} \label{eq:distinct}
D_n \sim \frac{a_D n}{logn},
\end{equation}
where $a_R$ and $a_D$ are constant. By fitting and comparing our
outcomes relevant to spin-1/2 and spin-1 systems
(Fig.~\ref{fig:Origin_Distinct_Short}), we find that
Eqs.~\ref{eq:origin} and  \ref{eq:distinct} formally still hold,
but $a_R$ and $a_D$ are functions of the temperature. More
precisely, in $T_c$ they show a maximum and a minimum,
respectively; such effect is more important when $S=1$
(Fig.~\ref{fig:Fit_Distinct_Short}) .

\begin{figure}
\resizebox{0.6\columnwidth}{!}{\includegraphics{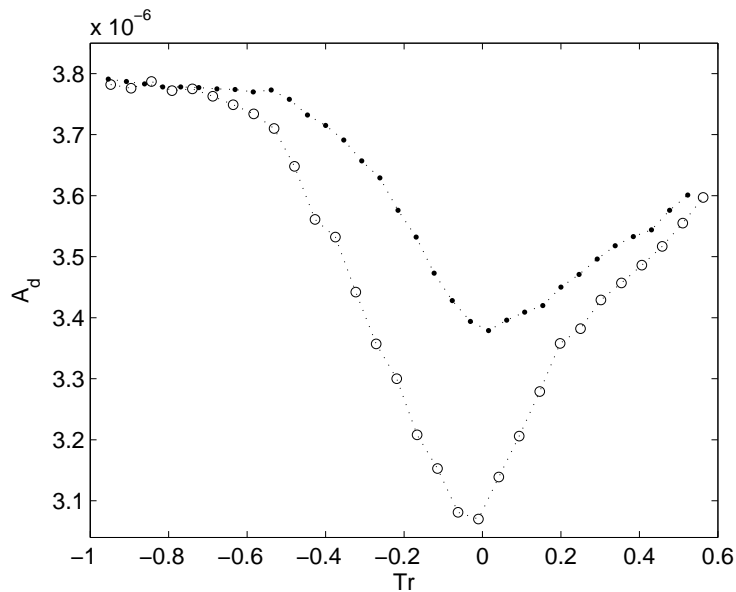}}
\caption{\label{fig:Fit_Distinct_Long} Fit parameter $A_D$ versus
reduced temperature pertaining to the BRW on a $240 \times 240$
spin-1/2 and spin-1 Ising lattice. Note that the minimum occurs at
about the critical temperature ($T_c \approx 2.6$ and $T_c \approx
1.96$ respectively) and it is deeper in the latter case.
Conversely, at low temperatures, plots are overlapped.}
\end{figure}

When the walk-length is large enough, for the walker, to
experience the toroidal effect, we find that the number of returns
$R_n$ recovers the URW case, being
\begin{equation} \label{eq:origin_long}
R_n \sim \frac{n}{N},
\end{equation}
while the number of distinct sites visited is consistent with
\begin{equation} \label{eq:distinct_long}
D_n \sim N \: (1 - e^{-nA_D}),
\end{equation}
relevant to the URW on a periodic lattice \cite{weiss2}, provided
that $A_D$ depends on $T$ (Fig.~\ref{fig:Fit_Distinct_Long})
However, by further increasing $n$, the bias effect vanishes also
for $D_n$ which, finally, equals $N$.

\begin{figure}[tb]
\resizebox{0.45\columnwidth}{!}{\includegraphics{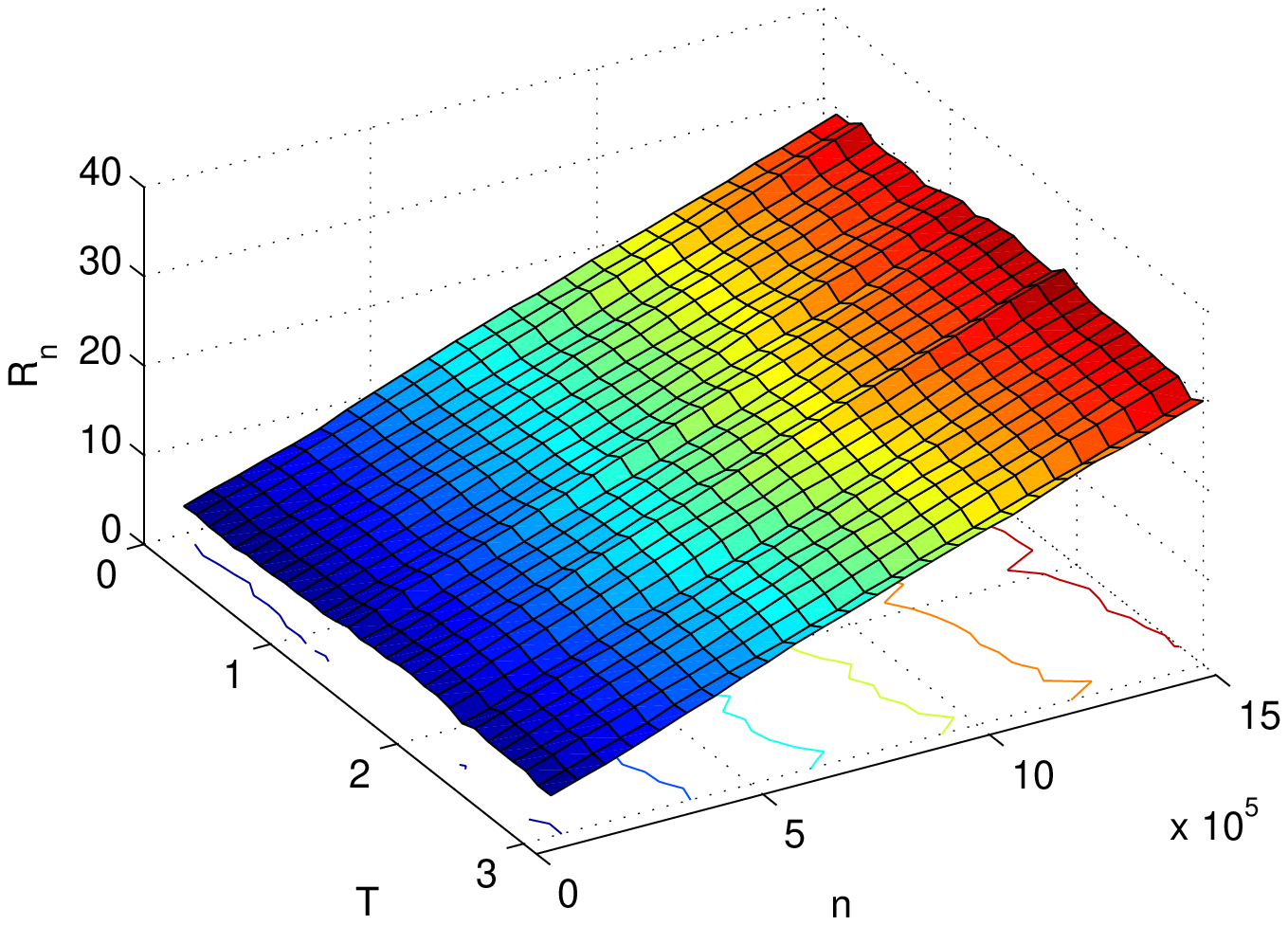}}
\resizebox{0.45\columnwidth}{!}{\includegraphics{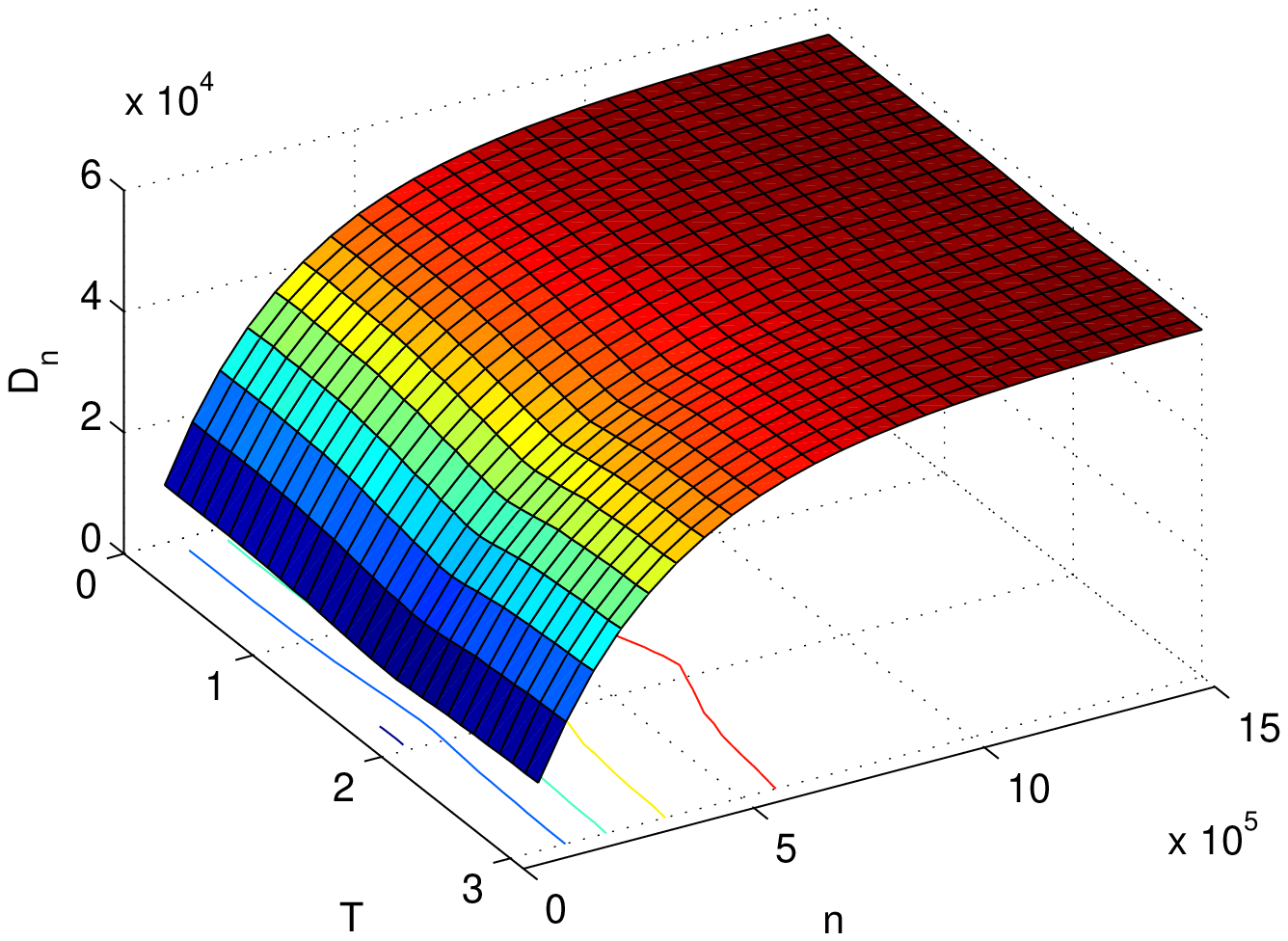}}
\caption{(Color online) \label{fig:Origin_Distinct_Long} Average
number of returns to the origin (left panel) and of distinct sites
visited (right panel) versus temperature and number of steps $n$
made by the BRW on a $240 \times 240$ spin-1 Ising lattice. In
this case $N \ll n$ and the walker can experience the periodic
boundary conditions the lattice is endowed with. Note that the
temperature dependence is now scarcely detectable. Similar results
were found for the spin-1/2 system.}
\end{figure}

Therefore, our results are asymptotically independent on the
temperature and on the spin magnitude, which means that the walker
looses memory of its bias.

It should be underlined that, in both regimes, the extreme points
recorded at $T_c$ for $R_n$, as well as $D_n$, are consistent with
what previously found about ${\cal T}_N$ and $\tilde{\epsilon}$.

Analogous results are expected also for the probability of return
to the origin $P_{0,n}$. In fact, we anticipate that the BRW still
recovers the unbiased exponent, with a multiplicative factor
maximum at $T_c$.

Notice that recovering the conventional diffusive regime from a
significantly altered model (at least in a long time limit) is
consistent with several previous works \cite{avramov,trimper}. The
fact that our model yields diffusive behavior should be related to
the absence of strong memory effects which, indeed, could
determine anomalous effects \cite{schulz1,schulz2}.

\section{Conclusions}
\label{concl} By the analysis performed so far, we are able to
characterize the BRW introduced and also to relate its behavior
with the evolution of the underlying energetic environment.

Our measures of local energy show that, according to the algorithm
introduced in Sec.~\ref{dyn}, the walker aims to move towards high
energy regions, where favorable spin-flips can occur. These
regions correspond to boundaries between clusters, whose
concentration depends on the temperature, being largest around
$T_c$. Thus, there exist a sort of temperature sensitive traps,
the walker is attracted to. This attraction reflects on a
non-homogeneous visit numbers distribution and also causes a
slowing down well evinced by measures of covering time, numbers of
returns to the origin and of distinct sites visited. In general,
these quantities asymptotically agree with results, analytically
known, relevant to the URW. In particular, when the walk-length is
not so large to visit all lattice sites, their functional forms
are URW-consistent, but temperature dependent multiplicative
factors have to be introduced. Their particular dependence on $T$
has two main consequences: first, the slowing down is not only due
to a non-null waiting probability, but it is mainly a consequence
of the bias introduced; second, the interaction between the walker
and the magnetic lattice is stronger in the critical region. More
precisely, in that temperature range, the length of borders
between clusters is large so that there are lots of high-energy
sites and, contemporary, the temperature is still not too large to
have a significant (with respect $\beta$) energy gain by
spin-flips. Hence, multiplicative factors just peak at the
critical temperature.

What has been said hitherto holds for either spin-1/2 and spin-1
systems. In fact, our analysis have been contemporary performed on
both, in order to evidence how the spin magnitude affects the
walker diffusion on the lattice. Then, our results show that peaks
get sharper and higher when the number of spin states is larger.
Actually, in the latter case, the walker has to manage with more
opportunities and, consequently, it is further slowed down.

In summary, in the limit $T \rightarrow 0$ ($T \rightarrow
\infty$) the underlying energetic landscape becomes homogeneous
and the BRW recovers the case of an ordinary unbiased random
walker endowed with a non-null waiting probability. On the other
hand, when $T=T_c$ we record the most important effects. In fact,
$T_c$ is an extremal point for the multiplicative factors which
correct the ordinary laws. Hence, the bias introduced determines
stronger effects as the critical temperature is approached, though
not affecting the diffusive regime. Therefore the BRW behavior
gives a further evidence of the phase transition.

Since such corrections are more important when the spin magnitude
is larger, we argue that an investigation in the continuum limit
for $S$ would turn out to be useful in order to clear the nature
of the above mentioned extreme points.


\begin{thebibliography}{}
%
\bibitem{avramov}
I. Avramov, A. Milchev and P. Argyrakis, Phys.\ Rev.\ E\
\textbf{47}, 2303 (1993).
%
\bibitem{bricmont}
J. Bricmont and A. Kupiainen, Phys.\ Rev.\ Lett.\ \textbf{66},
1689 (1991).
%
\bibitem{nitzan}
A. Nitzan and M. A. Ratner, J.\ Phys.\ Chem.\ \textbf{98}, 1765
(1994).
%
\bibitem{bustingorry}
S. Bustingorry, E. R. Reyes, and M. O. C\'{a}ceres, Phys.\ Rev.\ E
\textbf{62}, 7664 (2000).
%
\bibitem{harrison}
A. K. Harrison and R. Zwanzig, Phys.\ Rev.\ A \textbf{32}, 1072
(1985).
%
\bibitem{druger}
S.D. Druger, M.A. Ratner and A. Nitzan, Phys.\ Rev.\ B
\textbf{31}, 3939 (1985).
%
\bibitem{masuda}
N. Masuda and N. Konno, Phys.\ Rev. E \textbf{69}, (2004) 66113.
%
\bibitem{slutsky}
M. Slutsky, M. Kardar and L.A. Mirny, Phys.\ Rev.\ E\ \textbf{69},
61903 (2004).
%
\bibitem{bookout}
B.D. Bookout and P.E. Parris, Phys.\ Rev.\ Lett. \textbf{71}, 16
(1993).
%
\bibitem{levitsky}
I.A. Levitsky, Phys.\ Rev.\ B \textbf{49}, 15594 (1994).
%
\bibitem{arapaki}
E. Arapaki, P. Argyrakis, I. Avramov and A. Milchev, Phys.\ Rev.
E, \textbf{56}, 56 (1997).
%
\bibitem{haus}
W. Haus and K.W. Kehr, Phys.\ Rep. \textbf{150}, 263 (1987).
%
\bibitem{schulz1}
B. Schulz, S. Trimper and M. Schulz, Eur.\ Phys.\ J.\ B
\textbf{15}, 499 (2000).
%
\bibitem{schulz2}
B. Schulz, S. Trimper, Phys.\ Lett.\ A \textbf{256}, 266 (1999).
%
\bibitem{schulz3}
M. Schulz, S. Trimper, Phys.\ Rev.\ E \textbf{62}, 221 (2000).
%
\bibitem{sapag}
K. Sapag, V. Pereyra, J. L. Riccardo and G. Zgrablich, Surf.\
Sci.\ \textbf{295}, 433 (1993).
%
\bibitem{earlier}
E. Agliari, R. Burioni, D. Cassi and A. Vezzani, Eur.\ Phys.\ J.\
B, \textbf{46}, 109 (2005).
%
\bibitem{buonsante}
P. Buonsante, R. Burioni, D. Cassi and A. Vezzani, Phys.\ Rev. E,
\textbf{66}, 36121 (2002), and references therein.
%
\bibitem{montroll}
E.W. Montroll and G.H. Weiss, J.\ Math.\ Phys.\ \textbf{6}, 167
(1965).
%
\bibitem{weiss2}
G.H. Weiss, \textit {Aspects and Applications of the Random Walk}
(elsevier Science, Amsterdam 1994)
%
\bibitem{weiss}
H. Weiss and R.J. Rubin, Adv.\ Chem.\ Phys.\ \textbf{52}, 363
(1983).
%
\bibitem{burioni}
R. Burioni and D. Cassi, J.\ Phys.\ A \textbf{38}, (2005) R45.
%
\bibitem{jonasson}
J. Jonasson and O. Schramm, ArXiv Mathematics e-prints,
arXiv:math/0002034, (2000).
%
\bibitem{dembo} A. Dembo, Y. Peres, J. Rosen and O. Zeitouni, ArXiv Mathematics e-prints,
arXiv:math/0107191, (2001).
%
\bibitem{nemirovsky}
A.M. Nemirovsky, H.O. Martin and M.D. Coutinho-Filho, Phys.\ Rev.\
A \textbf{41}, 761 (1990).
%
\bibitem{trimper}
S. Trimper, U.C. T\"{a}uber, and G.M. Sch\"{u}tz, Phys.\ Rev.\ E
\textbf{62}, 6071 (2000).
%




\end{thebibliography}
\end{document}